\documentclass[runningheads]{llncs}

\usepackage{cite}
\usepackage{graphicx}
\usepackage{epstopdf}
\usepackage{array}
\usepackage{booktabs}
\usepackage{amsmath,amssymb} % define this before the line numbering.
\usepackage{color}
\usepackage{multicol}
\usepackage{multirow}
\begin{document}
\title{Stain Style Transfer using Transitive Adversarial Networks}
\author{Shaojin Cai\inst{1,2} \and
Yuyang Xue\inst{3} \and
Qinquan Gao\inst{1,2,3} \and
Min Du\inst{1,2}
Gang Chen\inst{4} \and
Hejun Zhang\inst{4} \and
*Tong Tong\inst{1,2,3}}
\authorrunning{S.J. Cai, T. Tong et al.}
\institute{College of Physics and Information Engineering, Fuzhou University, Fuzhou, China\and
Fujian Key Lab of Medical Instrumentation \& Pharmaceutical Technology, Fuzhou, China\and
Imperial Vision Technology, Fuzhou, China\and
Department of Pathology, Fujian Provincial Cancer Hospital, The Affiliated Hospital of Fujian Medical University, Fuzhou, China
%\email{*}\\
%\url{*}
%\and
%*\\
\\
\email{\{ttraveltong\}@gmail.com}}

\maketitle              % typeset the header of the contribution
\begin{abstract}
Digitized pathological diagnosis has been in increasing demand recently. It is well known that color information is critical to the automatic and visual analysis of pathological slides. However, the color variations due to various factors not only have negative impact on pathologist's diagnosis, but also will reduce the robustness of the algorithms. The factors that cause the color differences are not only in the process of making the slices, but also in the process of digitization. Different strategies have been proposed to alleviate the color variations. Most of such techniques rely on collecting color statistics to perform color matching across images and highly dependent on a reference template slide. Since the pathological slides between hospitals are usually unpaired, these methods do not yield good matching results. In this work, we propose a novel network that we refer to as Transitive Adversarial Networks (TAN) to transfer the color information among slides from different hospitals or centers. It is not necessary for an expert to pick a representative reference slide in the proposed TAN method. We compare the proposed method with the state-of-the-art methods quantitatively and qualitatively. Compared with the state-of-the-art methods, our method yields an improvement of 0.87dB in terms of PSNR, demonstrating the effectiveness of the proposed TAN method in stain style transfer.

\end{abstract}

\section{Introduction}
Staining is a general process in pathology. However, the differences in raw material, staining protocols and slide scanners between labs make the appearance of the pathological stain suffer from large variability. These variations not only affect the diagnosis of the pathologists \cite{one}, but also can hamper the performance of CAD systems \cite{two}.

As an alternative, algorithms for automated standardization of digitized whole-slide images (WSI) have been published \cite{three,four,five,six,seven,eight,night,ten}. These methods can be roughly divided into three categories. {\bfseries Color-matching based methods} that try to match the color spectrums between the image and reference template image. Reinhard et al. \cite{seven} matched the color-channels between the image and reference template image in the LAB color space. However, this global color mapping fails in some local regions of image, as the same transformation is applied across the whole image while ignoring the independent distributions of color in different areas of the pathological image.

In addition, {\bfseries Stain separation based methods} that do the normalized operations on each staining channel independently. Macenko et al. \cite{twelve} proposed the stain vectors by transforming the RGB to the Optical Density (OD) space. Khan et al. \cite{thirteen} assigned every pixel to the specific stain component and estimated the stain matrix. Bejnordi et al. \cite{fourteen} thought that these methods did not take the spatial features of the tissue into account, which might lead to improper staining. Moreover, picking a good reference image requires expert knowledge and a bad reference may hamper the performance of these methods. The third group are {\bfseries Deep-learning based approaches}. These methods take advantage of the Generative adversarial networks (GANs) to transfer stain style. BenTaieb \cite{fifteen} designed a stain normalization net based on GANs with a discriminative image analysis model on top. However, this stain style transfer model depends on a specific model for a specific task on top. Shaban \cite{sixteen} proposed a method which is known as StainGAN. StainGAN is based on an Unpaired Image-to-Image Translation using Cycle-Consistent Adversarial Networks (CycleGAN). Cycle-consistency allows the images to be mapped into different color models but preserving the same tissue structure.

In this paper, we propose Transitive Adversarial Networks (TAN). TAN is also based on an Unpaired Image-to-Image Translation using Cycle-Consistent Adversarial Networks (CycleGAN) \cite{seventeen}. We proposed a novel generator, which can result in more accurate color transfer than other generators. TAN not only eliminates the problem of picking the reference template image but also achieve much higher quality and much faster processing speed than StainGAN, making it easier to minish the stain variants and improving the diagnosis process of the pathologists and CAD system. We have compared our method with state-of-the-art methods quantitatively and qualitatively, which demonstrates superiority of the proposed method.
\section{Methodology}
\begin{figure}
\includegraphics[width=\textwidth]{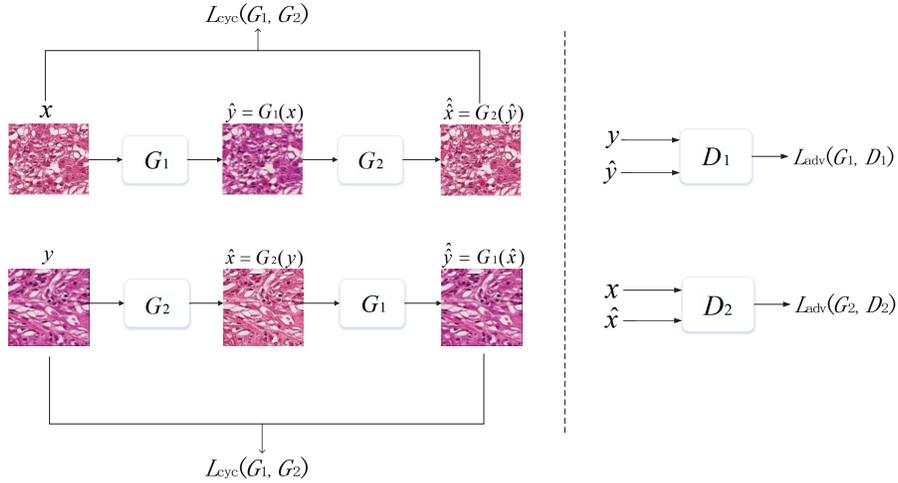}
\caption{The proposed framework for stain style transfer. $x$ and $y$ are unpaired images randomly sampled from their respective domains.} \label{fig1}
\end{figure}
\subsection{The Framework}
Our framework is illustrated in Fig.1. TAN is an unsupervised framework based on CycleGAN \cite{seventeen} in stain style transfer, which allows bidirectional transference of the H\&E Stain Appearance between different scanners, i.e from Aperio (A) to Hamamatsu (H) Scanner. This fremework does not require paired data from different scanners. The model consists of two generators $G_1 : A \rightarrow H$ and $G_2 : H \rightarrow A$. Each generator is trained with a corresponding discriminator, $D_1$ and $D_2$. For illustration, the first pair ($G_1$ and $D_1$), try to map images from domain A to domain H. The source images $x$ in the domain A is the input of the generator $G_1$, which yields generated images $\hat{y}$, $\hat{y} = G_1(x)$. Both the generated images $\hat{y}$ and the unpaired target images $y$ in the domain H are treated as inputs of the discriminator network $D_1$. During the training process, $G_1$ and $D_1$ compete with each other. $D_1$ acts as a binary classifier, trying to distinguish the generated images $\hat{y}$ and target domain images y. Due to the adversarial training process, $G_1$ tries to improve the quality of the generated images $\hat{y}$ to foolish $D_1$. This training producer is formulated as a min-max optimization which has a adversarial loss function :
\begin{equation}
L_{adv}(G_1,D_1)=E_{y \sim p_{data}(y)}[logD_1(y)]+E_{x \sim p_{data}(x)}[log(1-D_1(G_1(x)))]
\end{equation}
Analogous to the first pair of the generator network $G_1$ and the discriminator network $D_1$, the second pair ($G_2$ and $D_2$), try to map images from the domain H to the domain A, which replaces the input images as y and the output images as x. The training producer is also formulated as a min-max optimization process, and the loss function is $L_{adv}(G_2,D_2):$
\begin{equation}
L_{adv}(G_2,D_2)=E_{x \sim p_{data}(x)}[logD_2(x)]+E_{y \sim p_{data}(y)}[log(1-D_2(G_2(y)))]
\end{equation}
However, if the training process is merely guided by the adversarial loss, it may result in the non-convergence of the training process and lead to model collapse. Several images from source domain will map to the single image in the
\begin{figure}
\includegraphics[width=\textwidth]{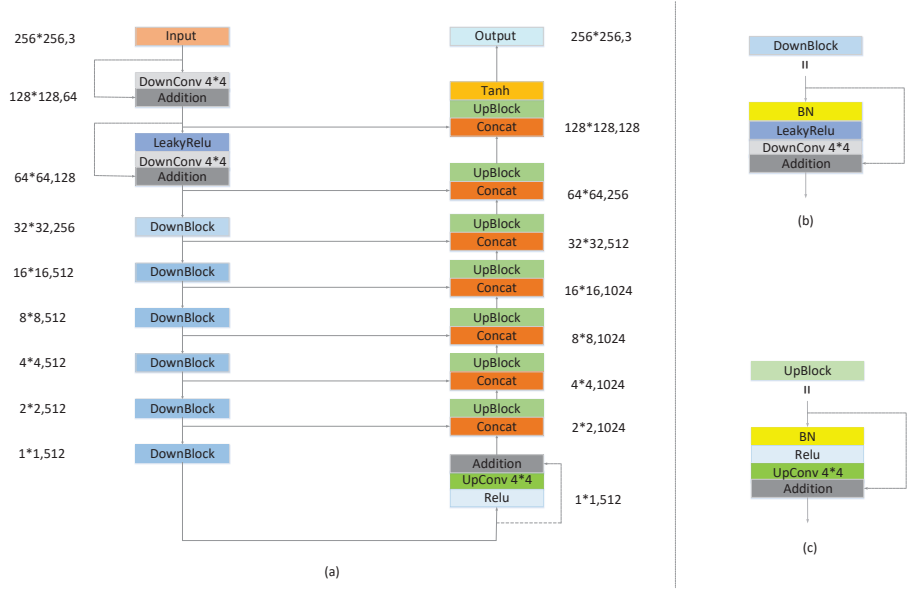}
\caption{The network of our proposed generator that we refer to as Trans-Net.} \label{fig2}
\end{figure}
target domain if only the adversarial loss is used. Therefore, additional training constraint on the mapping function is essential. This is achieved by adding a
cycle loss, which enforces the two mapping functions, $G_1$ and $G_2$ , to be cycle-consistent with each other. Generally speaking, two mapping functions should be
reciprocal, for illustration, $\hat{\hat{x}} = G_2(G_1(x))$, $\hat{\hat{y}} = G_1(G_2(y))$. This behaviour can be achieved by adding the pixel-wise cycle-consistency loss for both generators:
\begin{eqnarray}
L_{cyc}(G_1,G_2)&=E_{x \sim p_{data}(x)}[\Vert x-G_2(G_1(x)) \Vert _1 ]\nonumber\\
&+E_{y \sim p_{data}(y)}[\Vert y-G_1(G_2(y)) \Vert _1]
\end{eqnarray}
As a result, the final loss for the whole training process can be described as:
\begin{equation}
L(G_1,G_2,D_1,D_2)=L_{adv}(G_1,D_1)+L_{adv}(G_2,D_2)+\lambda L_{cyc}(G_1,G_2)
\end{equation}
\subsection{Network Architectures}
\subsubsection{\textbf{Generator}}Compared to U-Net, we have three innovations: i) We increase the numbers of downsampling layer and upsampling layer from 4 to 8 symmetrically, enabling the network to learn higher-level semantic information and generate more detailed context. And the experimental results demonstrate that we can produce the best images at the 8 level of downsampling. ii) As shown in Fig.2 (b) and Fig.2 (c), for each downsampling layer or upsampling layer, we add the information before sampling to the sampled information innovatively. The operation can propagate the information from previous layer to next layer, which takes advantage of low-level features with low complexity and high-level features with high complexity, making it easier to get a smooth decision function with better generalization performance and produce more detailed results both in color and texture. iii) In U-Net, convolution operations are performed twice before each sampling operation. But in Trans-Net, we delete the two convolution operations before each sampling layer. There is no convolution layer between adjacent sampling layers,  avoiding  the overfitting problem. And the network only includes 16 convolution layers  which is 7 layers less than U-Net, making it present a better image quality and less computation time.

The proposed generator architecture we refer to as Trans-Net is shown in Fig.2 (a). It consists of a coding operation (left side) and a decoding operation (right side). The sizes and numbers of the feature channels in every layer are written in the side of both paths. Shortcuts are used to concatenate the features from the coding phase to the decoding phase in all corresponding downsampling and upsampling blocks as shown in Fig.2 (a). This can avoid the gradient vanishing or exploding problem during backpropagation, making it easy to train deep networks.
\subsubsection{\textbf{Discriminator}}
Since the $L_1$ or $L_2$ term can successfully capture the low-frequency information but fail to restore high-frequency information, producing blurred details on image generation tasks \cite{twentyfive}. In order to generate both the low-frequency and the high-frequency details, we added a patch-level classifier as discriminator as proposed in \cite{twentyfour}. This discriminator we refer to as PatchGAN can learn high-frequency features while the $L_1$ loss can learn low-frequency features. By fusing the two types of losses, both the high-frequency and the low-frequency details can be learnt and generated. PatchGAN restricts the attention to the structure in local image patches, which penalizes the structure at the scale of $70 \times 70$, aiming to classify whether $70 \times 70$ overlapping image patches are real or fake. The output of PatchGAN is a $30 \times 30$ matrix where every element have a receptive filed of $70 \times 70$, averaging the matrix to provide the final output.

Although patch is much smaller than the image, they can still generate high quality results with fewer parameters, and faster inference speed than that at image level. In addition, it can work on arbitrarily-sized images in a fully convolutional fashion \cite{twentythree}. Such a discriminator effectively models the image as a Markov random field, assuming independence between pixels separated by more than a patch diameter. The Markov random field characterize the image by local fragmentation region of pixel values. Therefore, our PatchGAN can be treated as a form of texture/style loss.

\section{\textbf{Experiments and Results}}
To have fair and comprehensive comparisons with other methods, we evaluated our model as follows: i) Analysis of the image quality at different levels of downsampling;
\begin{figure}
\includegraphics[width=\textwidth]{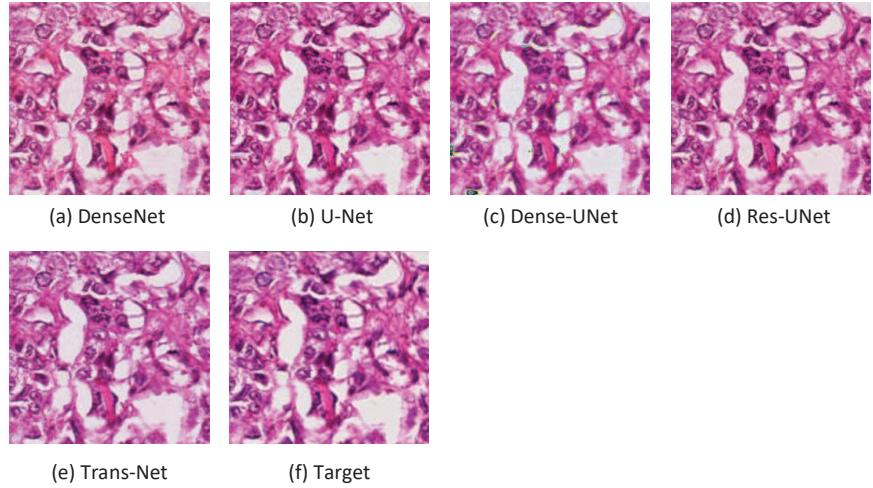}
\caption{Visual comparisons of results using different generators.} \label{fig3}
\end{figure}
ii) Analysis of the effect of the proposed generator on the results and comparisons of results using different generators; iii) Quantitative and qualitative comparisons between our method and state-of-the-art approaches \cite{sixteen}.
We will introduce the experimental dataset, the training details, the evaluational metrics and experimental results in the following sections.
\subsection{\textbf{Dateset and Details}}
\subsubsection{\textbf{Dataset}} The dataset is publicly available as part of the MITOS-ATYPIA14 challenge\footnote{https://mitos-atypia-14.grand-challenge.org}. The dataset consists of 424 frames at $X20$ magnification which were stained with standard Hematoxylin and Eosin (H\&E). The training dataset consists of 300 frames and the test dataset consists of 124 frames. All frames were scanned by two scanners: Aperio Scanscope XT and Hamamatsu Nanozoomer 2.0-HT. Slides from both scanners were resized to identical dimensions ($1539\times1376$). For training, we extracted 9000 unpaired patches from the training dataset of both scanners. During evaluation, we randomly extracted 620 paired patches from the testing dataset of both scanners. All patches have the same size of $256\times256$. Non-rigid registration was employed to eliminate the mismatch. Patches from Scanner H were regarded as the ground truth.
\subsubsection{\textbf{Training details}}
For all experiments, we trained 9000 unpaired patches from both scanners for 6 epoches with a batch size of 1. We used the Adam solver and trained all networks with a learning rate of 0.0002. We set $\lambda=10$ in Equation 4.
\begin{figure}
\includegraphics[width=\textwidth]{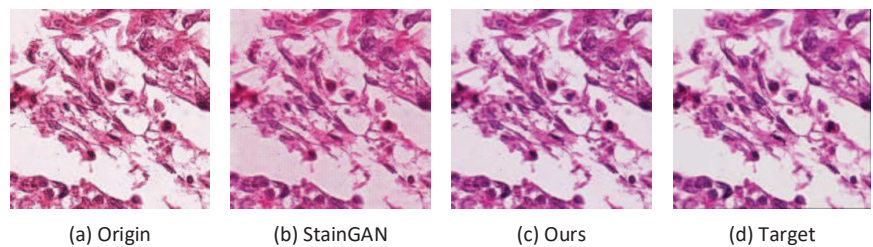}
\caption{Visual comparison between the result of our proposed method and that of StainGAN.} \label{fig4}
\end{figure}
\begin{table}
\caption{Results of Trans-Net at different levels : Mean indicators and total processing time}\label{tab1}
\begin{center}
\begin{tabular}{p{2.5cm}<{\centering}p{2.5cm}<{\centering}p{2.5cm}<{\centering}p{2.5cm}<{\centering}}
\hline
\specialrule{0em}{2pt}{2pt}
Methods & PSNR & SSIM & Time (sec) \\
\hline
\specialrule{0em}{2pt}{2pt}
Level8 &  22.22 & 0.812 & 31 \\
\specialrule{0em}{2pt}{2pt}
Level7 &  21.67 & 0.793 & 29 \\
Level6 & 21.99  & 0.807 & 28 \\
Level5 & 21.97  & 0.806 & 24 \\
Level4 & 21.99  & 0.802 & 22 \\
\hline
\end{tabular}
\end{center}
\end{table}
We replaced the negative log likelihood objective for $L_{GAN}$ by a least-squares loss \cite{twenty} and updated the discriminators using a history of generated images rather than the ones produced by the latest generators.
We kept an image buffer that stores the 50 previously created images. We used this image buffer which has information of previous 50 images rather than the latest image to update the discriminators. The hardware of GeForce GTX 1080 and the PyTorch framework were used.

\subsubsection{\textbf{Evaluation Metrics}} Results were compared to the ground truth with two similarity metrics: Peak Signal-to-Noise Ration (PSNR) and Structural Similarity index (SSIM). In addition, the processing speed which is an important factor in clinical has been reported in the results.
We used the total time over processing the 620 images from testing dataset to calculate the computational time.
\subsection{Results with different levels of downsampling}
It is interesting to see some results demonstrate the image quality with different levels of downsampling. We increase the numbers of downsampling layer and upsampling layer from 4 to 8 symmetrically. Results is shown in Table 1. It is obviously that the result at sampling level of 8 is the best. And the 8 level is the highest level that we can sample, \begin{table}
\caption{Comparison to other generators: Mean indicators and total processing time}\label{tab2}
\begin{center}
\begin{tabular}{p{2.5cm}<{\centering}p{2.5cm}<{\centering}p{2.5cm}<{\centering}p{2.5cm}<{\centering}}
\hline
\specialrule{0em}{2pt}{2pt}
Methods & PSNR & SSIM & Time (sec) \\
\hline
\specialrule{0em}{2pt}{2pt}
DenseNet & 21.40 & 0.792 & 35 \\
\specialrule{0em}{2pt}{2pt}
U-Net &  21.95 & 0.802 & 45 \\
\specialrule{0em}{2pt}{2pt}
Dense-UNet & 20.10 & 0.795 & 119 \\
\specialrule{0em}{2pt}{2pt}
Res-UNet &  22.02 & 0.802 & 51 \\
\specialrule{0em}{2pt}{2pt}
Trans-Net &  {\bfseries 22.22} & {\bfseries 0.812} & {\bfseries 31} \\
\hline
\end{tabular}
\end{center}
\end{table}
\begin{table}
\caption{Stain Transfer Comparison: Mean indicators and total processing time}\label{tab1}
\begin{center}
\begin{tabular}{p{2.5cm}<{\centering}p{2.5cm}<{\centering}p{2.5cm}<{\centering}p{2.5cm}<{\centering}}
\hline
\specialrule{0em}{2pt}{2pt}
Methods & PSNR & SSIM & Time (sec) \\
\hline
\specialrule{0em}{2pt}{2pt}
StainGAN &  21.35 & 0.785 & 60 \\
\specialrule{0em}{2pt}{2pt}
Ours &  {\bfseries 22.22} & {\bfseries 0.812} & {\bfseries 31} \\
\hline
\end{tabular}
\end{center}
\end{table}
because the size of our input is 256*256 and the size of the feature maps at this moment is 1*1. Then, we adopt 8 downsampling layers and 8 upsampling layers for our generative networks.
\subsection{Comparisons of results using different generators}
The generator of our generative adversarial networks we refer to as Trans-Net is based on the traditional U-Net structure \cite{eighteen}. In order to make a extensive comparisons, we replaced the generator with other network structures and compared the results using our proposed Trans-Net and other generators.

Four other network structures were used as generators for comparison, including the traditional U-Net \cite{eighteen} which was used in biomedical image segmentation, Res-UNet \cite{twentytwo} which was first proposed for segmentation of retinal images, the DenseNet structure which was used in classification tasks \cite{DenseNet}, and the Dense-UNet \cite{DenseUNet} which was used to remove artifacts from the image respectively. Results are shown in Table 2. The proposed Trans-Net achieved higher values than other generators in terms of PSNR and SSIM with less computational times. The visual comparisons as shown in Fig.3, demonstrates that the results generated by the proposed Trans-Net are closer to the ground truth than the results using other generators in terms of colors, contrast and texture details.
\subsection{Comparison with state-of-the-art method}
The goal is to transfer the style of pathes from scanner A (Aperio) to the patches from scanner H (Hamamatsu) while keeping the context and texture
of A. Since the medical images are unpaired in different centers, we used the cycle-consistent loss \cite{seventeen} to map the patches from domain A to domain H, and compared the generated patches with the real patches of scanner H (ground truth). The state-of-the-art method is StainGAN \cite{sixteen}, the difference between TAN and StainGAN is that we have designed a novel generator we refer to as Trans-Net. The author of StainGAN adopted the architecture for their generative networks from Johnson et al. \cite{twentythree} who have shown impressive results for neural style transfer and super-resolution. This network contains two stride-2 convolutions, 6 residual blocks, and two  stride-$\frac{1}{2}$ convolutions. Compared to the network, Trans-Net has much more stride-2 convolutions and stride-$\frac{1}{2}$ convolutions which results in more detailed semantic context information. And Trans-Net can produce much more detailed texture and color information owning to the operation that add the information before sampling to sampled information novelty. Finally, Trans-Net has only 16 convolutions, the reduction of convolutions accelerates the training speed and avoids the over-fitting problem. The results of the proposed method and StainGAN are shown in Table 3 , and Fig.4 shows their visual comparison. The PSNR value improves from 21.35 to 22.22 while SSIM improves from 0.785 to 0.812. In addition, the proposed method only requires half of the computational time of StainGAN. The visual comparison also show that the results generated by the proposed method are closer to the ground truth than the results using StainGAN.
\section{Discussion and Conclusion}
In this work, we proposed a novel method called TAN for stain style transfer. The experimental results show that the proposed method outperforms the state-of-the-art method in terms of objective metrics and visual comparisons. A new network structure called Trans-Net was proposed as generator, which contributes to better results than state-of-the-art results. There are three factors that contribute to the advantage of the Trans-Net structure: 1) It has many downsampling layers and upsampling layers to ensure the high-level semantic information can be learnt which can result in detailed texture and context. 2) It directly adds the information before and after sampling to reduce the loss of information which contributes to the much closer color and texture to the ground truth. 3) It only has 16 convolutional layers which accelerates the networks. It would be interesting to investigate how the stain style transfer affects the segmentation task and the analysis of pathological slides in our future work.
\section*{Acknowledgement}
This study was supported by the National Natural Science Foundation of China (Grant No. 6190010435) and the Science and Technology Program of Fujian Province, China (Grant No. 2019YZ016006).
\bibliographystyle{splncs04}
\bibliography{ebib}
\end{document}